\documentstyle[preprint,revtex]{aps}
\begin{document}
\draft
\preprint{UIOWA-94-01}
\def\to{\rightarrow}
\def\as{\alpha_s}
\def\qt{q_T}
\def\d{{\rm d}}
\def\tildeW{\widetilde{W}}
\def\VEV#1{\left\langle #1 \right\rangle}
\def\lsim{\buildrel <\over\sim}
\def\gsim{\buildrel >\over\sim}
\def\rmgev{{\rm GeV}}
\draft
\begin{title}
Relative distributions of W's and Z's at low transverse momenta\\
\end{title}
\author{M. H. Reno}
\begin{instit}
Department of Physics and Astronomy, The University of Iowa,\\
Iowa City, Iowa, 52242
\end{instit}
\begin{abstract}
Despite large uncertainties in the $W^\pm$ and $Z^0$
transverse momentum ($\qt$)
distributions for $\qt\lsim 10$ GeV, the ratio of the distributions
varys little. The uncertainty in the ratio of $W$ to $Z$ $q_T$ distributions
is on the order of a few percent, independent of the details
of the nonperturbative
parameterization.
\end{abstract}

\narrowtext

\section{Introduction}
\label{sec:intro}
With the precise measurements of the $Z^0$ mass from LEP experiments,
as well as other electroweak parameter measurements, the standard
model of electromagnetic and weak interactions is being
tested beyond tree level. Even without knowing the
masses for the top quark and the Higgs boson,
physical observables when compared to theoretical predictions
constrain new physics.\cite{lang}
An improvement in the measurement of the $W$ mass,
currently with a value of $80.22\pm 0.26$ GeV\cite{pdg} from CERN\cite{ua2}
and Fermilab\cite{cdf} Collider experiments,
would even further those
constraints.
With the higher statistics expected in the 1993-1994 Fermilab
Tevatron Collider runs, the theoretical
and systematic errors must be also be reduced to make a significant
improvement in the $W$ mass measurement.
Even though the dependence on the
$W$ transverse momentum ($q_T$) is not very strong
in the $W$ mass sample,
the QCD predictions for the low $\qt$ production of
$W$'s, including nonperturbative effects, must be better understood.

The low $\qt$ behavior of $W$'s and $Z$'s has been the subject of much
theoretical work. The Collins-Soper-Sterman formalism\cite{css}
for the low transverse momentum distribution in Drell-Yan has been applied
to $W$'s and $Z$'s by Davies, Stirling and Webber,\cite{dsw}
Altarelli {\it et al.},\cite{alt}
and the matching of low and high $q_T$ by Arnold and
Kauffman.\cite{ak} A study of the parameterization of
nonperturbative effects in the low $\qt$ region has
recently been done by Ladinsky and Yuan.\cite{cp}
Theoretical uncertainties in the separate $W$ and $Z$
$\qt$ distributions are discussed in detail in Ref.~\cite{ak}.
Here, we discuss
the {\it relative} uncertainties between the $W$ and $Z$ transverse momentum
distributions.
By considering the ratios
\begin{equation}
R(\qt ,y )\equiv\Bigl( {\d\sigma \over \d\qt\d y}(W^\pm)\Bigr)
 /  \Bigl( {\d\sigma \over
\d\qt\d y}(Z^0)\Bigr)\, ,
\end{equation}
and
\begin{equation}
R(\qt  )\equiv \Bigl( {\d\sigma \over \d\qt}(W^\pm) \Bigr) /
\Bigl(  {\d\sigma \over
\d\qt}(Z^0)\Bigr)
\end{equation}
the theoretical errors are reduced because the uncertainties
tend to cancel.
By comparing these quantities with experimental results, one can test
the theoretical treatment of low $\qt$ vector bosons. Alternatively,
the ratios can be used in Monte Carlo models for the $W$ mass measurement.

The estimation of the theoretical uncertainties in $R(\qt ,y)$ and
$R(\qt )$ due to the parameterization of nonperturbative
effects is the main topic of this paper.
In Section 2, we review the theoretical issues associated with the
low transverse momentum behavior of $W$'s and $Z$'s. Our numerical
results are displayed in Section 3. Section 4 contains our conclusions.
As all of the relevant formulae are collected
in the Appendix of Ref.~\cite{ak}, we do not include detailed formulae here.

\section{Theoretical Issues}
\label{sec:theo}

The calculation of the low transverse momentum behavior of $W$'s and
$Z$'s is not completely straightforward because of the presence of large
logarithms associated with powers of $\alpha_s$.
At low $\qt$, the perturbative expansion in
$\as$ is ruined because of the accompanying large logarithms. However,
it is possible to resum the leading and next-to-leading $\ln (Q^2/\qt^2)$
powers for each power of $\as$ to get an exponential factor.
Collins, Soper and Sterman have done the resummation,\cite{css} explicitly
including transverse momentum conservation for multiple parton
emission at each order in $\as$. The momentum conservation appears
through a Fourier transform involving the conjugate variable
to $\qt$, namely the
impact parameter $b$.
The resummation appears as a Sudakov-like exponent $S$, which is incorporated
in a form factor, here called $\tildeW$.
The quantity $\tildeW$ can be computed perturbatively over the
range of $\Lambda_{QCD}\ll 1/b<Q$.
Schematically
\begin{equation}
\tildeW_{{\rm pert}}(b,Q,x_A,x_B)\sim \sum_{ij}e_{ij}^2(V)
f_{i/A}(x_A,\mu )f_{j/B}(x_B,\mu )\exp [-S_{{\rm pert}}(b,Q )] \, .
\end{equation}
The parton distribution
functions $f_{i/A}$ are labeled
for parton $i$ in hadron $A$. The full
perturbative form for $\tildeW$
also includes convolutions of the parton distribution functions
with other functions which are suppressed by $\as /(2\pi )$.
The coupling constants $e_{ij}(V)$ depend on whether $V=W$ or $V=Z$.

Including a normalization parameter
$N_V,\ V=W$ or $Z$,
the differential cross section has the form
\begin{equation}
{\d\sigma\over \d\qt^2\d y}={\pi \alpha\over 3 s}N_V
\int_0^\infty \d^2b\, \exp (i\vec{\qt}\cdot\vec{b} )
\cdot
 \tildeW (b,Q,x_A,x_B,\mu )
\end{equation}
in the low $\qt$ limit.
Eqn. 4 applies only to the low $\qt$ limit because not included is the
part of the cross section which is regular as $\qt\rightarrow 0$.
Arnold and Kauffman have shown that the regular part of the differential
cross section for $q_T^W$ and $\qt^Z$ has little effect below
$\qt = 10$ GeV.
As we shall see below, the nonperturbative contributions are sizable only
for $\qt\lsim 10$ GeV,
so we consider this region of low $\qt$ and neglect
the regular part throughout.
The form factor $\tildeW$ depends on the impact parameter, the
mass of the weak gauge boson
and on
$x_A=(Q/\sqrt{ S})e^y$
and $x_B=(Q/\sqrt{ S})e^{-y}$, where $\sqrt{S}$ denotes the hadron
collider center of mass energy, and $y$ is the weak gauge boson
rapidity.
The form factor also includes the dependence on the parton distribution
functions, the resummed terms, a nonperturbative factor
and the unphysical factorization scale $\mu$.

For $b\leq 1/Q$, we roughly approximate $\tildeW$,
since in the integral in eqn. 4, this range of $b$ contributes little
for $\qt\lsim 10$ GeV.
However, for $b\gsim 1/\Lambda_{QCD}$, the perturbative expression
for $\tildeW$ breaks down. A cutoff in $b$ is introduced
in the standard way by\cite{css,dsw,ak}
\begin{equation}
b\rightarrow b_\star={b\over \bigr( 1+b^2/b_{max}^2\bigl)^{1/2} }
\end{equation}
where $b_{max}\sim 1/Q_0$ is
characterized by a scale $Q_0$ at which perturbation theory
is still valid.
The substitution of $b\to b_\star$ in
$\tildeW_{{\rm pert}}$ is accompanied
by a nonperturbative factor.
Thus, $\tildeW$ defined over the full range of $b$ is
written this way:
\begin{equation}
\tildeW(b,Q,x_A,x_B,\mu )=\tildeW_{{\rm pert}}(b_\star ,Q,x_A,x_B,\mu )
\cdot
e^{-S_{np}(b,Q)}\ .
\end{equation}
By combining eqns. (3)-(6),
the main theoretical uncertainties are evident:
dependence on the parton distributions and $\Lambda_{QCD}$, the
residual $\mu$ dependence, and the parameterization of the
nonperturbative effects in $S_{np}(b,Q)$. The nonperturbative
effects dominate for $\qt\lsim 6$ GeV, so we focus
our attention on the nonperturbative parameterizations.

Not indicated in the formulae above is an additional dependence on
three unphysical constants denoted $C_1,\ C_2$ and $C_3$ in Ref. \cite{ak}.
Arnold and Kauffman showed that the dependence of $\d\sigma /\d\qt$
for $W$'s and $Z$'s on these three parameters is weak for a
range of parameters.\cite{ak}

The nonperturbative factor $S_{np}$ has the form:\cite{css}
\begin{equation}
S_{np}(b,Q)=G_1(x_A,x_B,b,b_{max})
+G_2(b,b_{max})\ln (Q b_{max})
\ .
\end{equation}
In principle, $G_1$ depends on the incoming parton type and is hadron
dependent. Following
Davies, Stirling and Webber\cite{dsw}, we
will assume that $G_1$ is independent of parton type and
independent of $x_A$ and $x_B$. Ladinsky and Yuan include a dependence
on $x_A$ and $x_B$ in their parameterization,\cite{cp} which we comment
on below.
The standard assumption is that
$G_1$ and $G_2$ are proportional to $b^2$,\cite{css,dsw}
and $S_{np}(b,Q)$ is written in terms of constants $g_1$ and $g_2$:
\begin{equation}
S_{np}(b,Q)=(g_1+g_2\ln(Qb_{max}/2))b^2\ .
\end{equation}
This is, at large $b$,
roughly equivalent to a Gaussian distribution of intrinsic momentum.
Ignoring the $b$ dependence in $\tildeW_{{\rm pert}}$, one is left with the
Fourier transform of a Gaussian, which is itself a Gaussian.
The average
transverse momentum squared in the Gaussian is
$\VEV{\qt^2}=4[g_1+g_2\ln(Q b_{max}/2)] $.

Davies, Stirling and Webber\cite{dsw} have performed a fit to ISR (R209)
and Fermilab (E288)
fixed target data. Their numerical values for $g_1$ and $g_2$ are:
\begin{eqnarray}
g_1 &=&0.30 \ \rmgev^2\quad\quad g_2=0.16\ \rmgev^2 ~~ (np1)\nonumber \\
g_1 &=&0.15\ \rmgev^2 \quad\quad g_2=0.41\ \rmgev^2 ~~ (np2) \nonumber\\
          g_1& =& 0.0\ \ \rmgev^2\quad\quad
g_2=0.60\ \rmgev^2  ~~ (np3)\nonumber
\end{eqnarray}
with $b_{max}=0.5$ GeV$^{-1}$. The nonperturbative parameterization
$np2$ is their preferred fit. These parameters
were used by Arnold and Kauffman in their analysis of theoretical
errors in $\d\sigma /\d \qt$.

Evident from the discussion so far is the fact that the
the distributions for $Q=M_W$ and $Q=M_Z$ will not be very much
different apart from the overall normalization factors
and the combination of parton distribution functions, evaluated
at slightly different values of $x_A$ and $x_B$.
We now proceed to consider the numerical results.

\section{Numerical Results}
\label{sec:nume}

In obtaining the numerical results, we have used
$M_W=80$ GeV and $M_Z=91$ GeV. The weak mixing angle was taken such that
$\sin^2\theta_W =0.22$.
To compare nonperturbative parameterizations, we have used the Harriman
{\it et al.}\cite{hmrs} parton distribution functions
HMRSB with the four flavor
$\overline{{\rm MS}}$ scale $\Lambda_{QCD}=0.19$ GeV, based on a fit
to the BCDMS data.
To compare the effect of a change in parton distribution functions
on $R(\qt ,y=0 )$, we also use the HMRSE distribution functions fit
to the EMC data with $\Lambda_{QCD}=0.10$ GeV. As in Ref. \cite{ak}, the
canonical choices of $C_1=C_3=2 e^{-\gamma_E}\equiv b_0$ and $C_2=1$
are used. Here, $\gamma_E$ is the Euler constant.
As indicated above, we neglect the regular part of the differential
cross sections, which contribute on the order of less than 1\%  in
the range of $\qt<10$ GeV.\cite{ak}
All of the figures show results for
the Tevatron Collider at $\sqrt{S}
=1.8$ TeV. Similar results obtain for the CERN Collider.

\subsection{Nonperturbative parameters $g_1$ and $g_2$}
\label{subsec:nonp}

The largest uncertainty for $\qt\lsim 6$ GeV comes from the choice of
parameters $g_1$ and $g_2$ in eqn. (8).
Figure 1 shows the distribution $\d\sigma /(\d\qt\d y)$
at $y=0$ as a function of $\qt$ of the $W$, including both charges of the
$W$ in the rate. The three predictions vary from the average of the three
curves by as much as 40\% for very low $\qt$. For
$\qt\gsim 6$ GeV, the the sensitivity to the nonperturbative parameterization
is significantly less. Unfortunately, it is precisely below
$\sim 6$ GeV where the differential cross section is the largest.
A similar uncertainty is obtained for the $Z^0$ transverse momentum
distribution as a function of the nonperturbative parameterization.

Because the $W$ and $Z$ masses are close and there is only log dependence
on the masses,
the nonperturbative parameterization $S_{np}(b,Q)$ and
perturbative exponent $S_{{\rm pert}}(b_\star, Q)$ are nearly identical
for $Q=M_W$ and $Q=M_Z$. One expects that the ratio of the $W$ to
$Z$ transverse momentum distributions will be approximately constant
for small $\qt$. Indeed, this is the case.
Figure 2
shows the relative size of the $W$ to $Z$ distributions at
$y=0$ for $np1-np3$.
Qualitatively, the size of $R(\qt,y=0)$ can be understood from a combination of
coupling constants and parton distribution functions.
Consider the quantity
\begin{equation}
D_V(x_A,x_B,\mu )=\sum_{ij} e_{ij}^2(V)f_{i/p}(x_A,\mu )f_{j/\bar{p}}
(x_B,\mu )\ \nonumber
\end{equation}
By taking
\begin{equation}
D_W(M_W/\sqrt{S},M_W/\sqrt{S}, M_W)/D_Z(M_Z/\sqrt{S},M_Z/\sqrt{S}, M_Z)\, ,
\nonumber
\end{equation}
one obtains the value of 3.04. If one were to evaluate the $W$ and $Z$ parton
luminosities at the same values of $x_A$ and $x_B$, the ratio would equal
2.34, as a consequence of their different coupling constants.
Deviations of the $np1-np3$ curves from the average value
$R_{{\rm avg}}(\qt ,y=0)$ are less than a percent.

To understand the increase in $R$ as $\qt$ decreases, it is
useful to look at $\tildeW $ for $W$'s and $Z$'s, which is independent
of $\qt$. In Figure 3$a$, we show the form factors
$\tildeW(b,Q,x_A,x_B,b_0/b_\star )$ with $x_A=x_B=Q/\sqrt{S}$
for $Q=M_W$ and $Q=M_Z$, and with the nonperturbative
parameters $np2$, as a function of $\ln (b\cdot {\rm GeV})$.
The low $b$ behavior is an extrapolation of the value of $\tildeW$
at $b=1/Q$, to lower values of $b$. The integral over $b$ of $\tildeW$
also includes a factor of $bJ_0(b\qt )$, so the differential distribution
is fairly
insensitive to such small values of $b$ where the extrapolation must
be performed.
For large values of $b$, the nonperturbative parameterization cuts off
the integral.
Figure 3$b$ shows the ratio of the form factors $\tildeW$, for the three
sets of nonperturbative parameters.
We cut off the range of $\ln (b\cdot {\rm GeV})$ displayed because
at low $b$, the shape of the ratio is an artifact of the
low $b$ extrapolation.
The curves for $np1-np3$
separate at $b\sim 10^{-1}$GeV$^{-1}$, because of the mass dependence
in $S_{np}$, with the largest ratio belonging to $np3$.

At low $\qt$, the
average value of $b$
is for ``large'' $b$. For example, for $W$'s
at $\qt=0.5$ GeV, $\VEV{b}=0.8$ GeV$^{-1}$.
As $\qt$ increases, the Bessel function
in the integral oscillates more rapidly as a function of $b$, and $\VEV{b}$
decreases, moving into the perturbative range.
Thus, as $\qt$ increases, the sensitivity to nonperturbative physics
decreases and the calculation is more reliable.
Figure 3$b$ also shows the trend
that $R(\qt )$ and $R(\qt ,y)$ decline with
increasing $\qt$.
If one trusts the form of the nonperturbative parameterization, then
the uncertainty in $R(\qt,y=0)$ is less than a few percent.
Even if one estimates
the uncertainty in $R$ by
the deviation from the ratio of coupling constants and parton
distribution functions, one is still left with a small
uncertainty due to nonperturbative parameterizations of order $5\%$.

Figure 4 shows the ratio $R(\qt )$ (where now the rapidity of the
gauge boson has been integrated) for $np1-np3$, as a function
of $\qt$. The shapes of the distributions are nearly identical to
those of Figure 2, however, with an increase in the normalization.
This is due to the fact that the ratio of the $W$ to $Z$ parton
luminosities increases with increasing $|y|$.
We have the same qualitative conclusions about the
uncertainty in the ratio $R(\qt )$ as in the $y=0$ case.

\subsection{The Ladinsky-Yuan parameterization}
\label{subsec:lyua}

The Ladinsky-Yuan parameterization\cite{cp}
is characterized by three parameters
and depends on $S$ as well as $Q$.
In a recent reanalysis of the R209 and E288 low $\qt$ Drell-Yan data,
Ladinsky and Yuan (LY) have postulated a modified form for $S_{np}$:
\begin{equation}
S_{np}(b,Q,x_A,x_B) = g_1b^2+g_1g_3b\ln (100 x_A x_B)
 +g_2b^2
\ln (Q b_{max}/1.6)
\end{equation}
The inclusion of the $x_A$ and $x_B$ dependence is to account for
changing $\VEV{\qt^2}$ as a function of $Q^2/S$.
The $W$ and $Z$ $\qt$ distributions have peaks at slightly higher
$\qt$ values than in the two parameter cases.
We use their central
values of $g_1,\ g_2$ and $g_3$ to characterize the effects of this
parameterization on $R(\qt ,y)$ and $R(\qt )$. Their central values are
$g_1=0.11\ \rmgev^2$, $g_2=0.58 \ \rmgev^2$ and $g_3=-1.5\
\rmgev^{-1}$, and their
fit was performed using the CTEQ parton distribution
functions.\cite{cteq}
As we are interested primarily in the ratios, we continue to use
the HMRSB parton distribution functions. The LY curve for $R(\qt )$
at $\sqrt{S}=1.8$ TeV is shown as the dotted line in Figure 4.
The deviation from the two parameter result is less than a few percent.

\subsection{Parton distribution dependence}
\label{subsec:part}

Figure 5 shows the ratio
of the $W$ and $Z$ $\qt$ distributions for two different parton
distribution functions.
We show
$R(\qt ,y=0 )$ for the HMRSB and HMRSE parton distribution
functions with $np2$.
The HMRSE distributions yield the slightly lower value of
$R$, but the two predictions for $R_{{\rm avg}}$ are within 2\% of
each other at fixed $\qt$.

\section{Conclusions}
\label{sec:conc}

The values of $g_1$ and $g_2$ determined by Davies {\it et al.}\cite{dsw}
are clearly not the best current values because of the advances in
our knowledge of the parton distribution functions. Ladinsky and
Yuan,\cite{cp} by
refitting the data with current parton distributions, have made
a significantly better choice. Nevertheless, uncertainties still exist
in the form of the parameterization, affecting the extrapolation from
$Q\sim 5$ GeV to $Q\sim 80-91$ GeV. Parton distribution uncertainties
at large $Q$ compound the uncertainties.\cite{cp}

We have shown that by taking the ratio of $W$ to $Z$ $\qt$ distributions,
the theoretical uncertainties due to the parameterization of nonperturbative
effects tend to cancel. For the range of $\qt<10$ GeV, the ratio
$R(\qt )$ is $\sim3.3$, slightly
higher or lower depending on $\qt$. When one
takes the ratio of the total cross sections obtained by integrating the
resummed, matched differential cross sections\cite{ak} for $np2$, one obtains
$R=3.27$. $R$ is less than 3.3 because
with increasing $\qt$, $R(\qt )$ declines
roughly linearly, with $R(\qt = 100$ GeV$)=2.66$.\cite{seep}

In our focus on the nonperturbative effects, we have neglected the regular
part of the differential cross section which comes in at the percent
level. A full study would also include an analysis of the other parameter
dependence described in the text, {\it e.g.}, $C_1$, $C_2$, $C_3$,
$b_{max}$, {\it etc.} We expect, however, that in the ratios
$R(\qt ,y)$ and $R(\qt )$ the dependence should largely cancel.

\acknowledgements
This work was initiated at the Aspen Center for Physics.
We thank H. Frisch for suggesting the comparison and
H. Frisch, K. Einsweiler and H. Baer for useful
conversations.
This research was
supported in part by NSF Grants PHY-9104773 and PHY-9307213.

\figure{The distribution $\d\sigma /(\d\qt\d y)$
at $y=0$ as a function of $\qt$ of the $W$ for the sum of $W^+$ and
$W^-$ production at $\sqrt{S}=1.8$ TeV,
with nonperturbative parameterizations {\it np1} (solid line),
{\it np2} (dashed line) and {\it np3} (dot-dashed line).
}
\figure{$R(\qt ,y=0)$ defined in eqn. 1 for $np1-np3$, as a function
of $\qt$ for $\sqrt{S}=1.8$ TeV. }
\figure{$a$) $\tildeW_{{\rm pert}} (b_\star,Q)\exp(-S_{{\rm np}}
(b,Q))$ with
$y=0,\ \sqrt{S}=1.8$ TeV and $\mu=b_0/b_\star$, as a function of
$\ln (b\, {\rm GeV})$. The solid line is for $Q=M_W$ and the dashed line
is for $Q=M_Z$.
$b$) Ratio of $\tildeW$ for $W^\pm$ to $\tildeW$ for $Z^0$ production,
as a function of $\ln (b\, {\rm GeV})$, for {\it np1} (solid line),
{\it np2} (dashed line) and {\it np3} (dot-dashed line).}
\figure{$R(\qt )$ defined in eqn. 2 for $np1-np3$, (solid, dashed and
dot-dashed lines) as a function
of $\qt$ for $\sqrt{S}=1.8$ TeV. The dotted line indicates $R(\qt )$ for
the Ladinsky-Yuan
parameterization.\cite{cp}}
\figure{$R(\qt ,y=0 )$ for the HMRSB (solid line) and
HMRSE (dashed line) parton distribution
functions for $\sqrt{S}=1.8$ TeV and $np2$.}

\end{document}